\documentstyle[12pt]{article}

\newcommand{\sgs}{\renewcommand{\baselinestretch}{1.0}
        \large\normalsize}
\begin{document}
\input epsf
\title{\bf The surfactant effect in semiconductor thin film growth}
\author{Daniel Kandel$^{(a)}$ and Efthimios Kaxiras$^{(b)}$\\
\vbox{\vspace*{8mm}}
$^{(a)}$ Department of Physics of Complex Systems \\ 
Weizmann Institute of Science\\
Rehovot 76100, Israel \\
\vbox{\vspace*{6mm}}
$^{(b)}$ Department of Physics and \\
Division of Engineering and Applied Sciences\\
Harvard University, Cambridge MA 02138, USA
}
\maketitle

\bigskip

\bigskip

I. Introduction \hfill \\ 

II. Experimental Observations \hfill \\ 
\hbox{\hspace{1cm}} 1. Group-IV films on group-IV substrates \hfill \\
\hbox{\hspace{1cm}} 2. III-V films on III-V substrates \hfill \\
\hbox{\hspace{1cm}} 3. Mixed film and substrate systems \hfill \\

III. Theoretical Models \hfill \\
\hbox{\hspace{1cm}} 1. Microscopic models \hfill \\ 
\hbox{\hspace{1cm}} 2. Macroscopic models \hfill \\ 

IV. The Diffusion--De-exchange--Passivation Model \hfill \\
\hbox{\hspace{1cm}} 1. General considerations \hfill \\ 
\hbox{\hspace{1cm}} 2. First-principles calculations \hfill \\ 
\hbox{\hspace{1cm}} 3. De-exchange and generalized diffusion \hfill \\ 
\hbox{\hspace{1cm}} 4. Island-edge passivation \hfill \\ 
\hbox{\hspace{1cm}} 5. Kinetic Monte Carlo simulations \hfill \\ 

V. Discussion \hfill \\

\bigskip


\section*{I. Introduction}

Progress in the fields of electronic and optical devices
relies on the ability of the semiconductor industry to 
fabricate components of ever increasing complexity and 
decreasing size.  The drive for miniaturization has actually  
provided the impetus for much fundamental and applied research 
in recent years. 
Nanoscale structures in 2, 1
or 0 dimensions, referred to as quantum wells,
wires and dots respectively, are at the forefront of exploratory
work for next generation devices.
In most cases these structures must be fabricated through 
epitaxial growth of semiconductor thin films, 
in either homoepitaxial (material A on substrate A) 
or heteroepitaxial (material A on substrate B) mode. 
There are usually 
two important requirements in this 
process:  First, the film must be of high quality crystalline material,
and second, a relatively low temperature must be maintained
during growth.
The need for the first requirement is self evident, since 
highly defected crystals typically perform poorly in electronic
applications; the imperfections, usually in the form of dislocations,
grain boundaries or point defects, act as electronic traps and degrade
the 
electronic properties to an unacceptable level.
The second requirement arises from the 
need to preserve the characteristics of the substrate during 
growth, such as doping profiles and sharp interfaces
between layers, which can be degraded due to atomic diffusion  
when the growth temperature is high.     
These two requirements seem to be incompatible: In order to 
improve crystal quality, atoms need to have sufficient surface 
mobility so that they can find the proper crystalline sites
to be incorporated into a defect-free crystal.  On the other hand, 
excessive atomic mobility in the bulk must be avoided, which can 
only be achieved by maintaining lower than typical growth 
temperatures. 
These problems are exacerbated in the case of heteroepitaxial 
growth, where the presence of strain makes smooth,
layer-by-layer growth problematic
even when the temperature is high, since the equilibrium
structure involves strain relieving defects. 

Early studies of the effects of contaminants and impurities on growth 
modes had indicated that it is indeed possible to alter the mode
of growth by the inclusion of certain elements, different than either
the growing film or the substrate.  
For a discussion of epitaxial growth 
modes and the effect of 
contaminants see the authoritative review by Kern et al. \cite{kern79}.  
A breakthrough in the quest for controlled growth
of semiconductor films was reported in 1989 when 
Copel et al.\ \cite{copel89}
demonstrated that the use of a single layer of As can improve the 
heteroepitaxial growth of Ge on Si, which is otherwise difficult due to
the
presence of strain.  Growth in this system typically proceeds
in the Stranski-Krastanov mode, that is, it begins with a few 
(approximately 3)
wetting layers but quickly reverts to three-dimensional (3D) 
island growth.
The eventual coalescence of the islands unavoidably
produces
highly defective material.  
In the experiments of Copel et al., 
the monolayer of As was first deposited to the Si substrate,
and continued to float during the growth of the Ge overlayers.
The presence of the As monolayer 
in the system led to a drastic change in both the 
thermodynamics (balance of surface and interface energies) and
the kinetics (surface and bulk mobility of deposited atoms), 
making it possible
to grow a Ge film in a layer-by-layer fashion, to unprecedented
thicknesses for this system (several tens of layers).
This remarkable behavior was termed the ``surfactant effect''
in semiconductor growth.  Since then, a large number of 
experiments has confirmed this behavior in a variety of 
semiconductor systems (for early reviews see 
\cite{tournie93c,ploog93,tournie94b}).  

We digress momentarily to justify the terminology.
The typical meaning of the word surfactant is 
different than here,  
which has led to some debate about the appropriateness
of the term in the context of semiconductor thin film growth.
The word surfactant,  
as defined in 
scientific dictionaries, 
is used commonly
in chemistry to describe
{\em ``a substance that lowers the surface or interfacial tension 
of the medium in which it is dissolved''}\ \cite{schram}, 
or {\em ``a material that improves the emulsifying, dispersing, 
wetting, or other surface modifying properties of liquids''}\
\cite{nasa}.
While these physical situations and the effect itself are 
different than the systems considered in the present review, 
we adopt here the term ``surfactant'' to describe the 
effect of adsorbate layers in semiconductor thin film growth
for two reasons. 
First, for a reason of substance: there are 
indeed some similarities between adsorbate layers 
in semiconductor thin film growth and the classical
systems to which the term applies; namely, in both cases 
the presence of this extra layer reduces the surface tension
and changes the kinetics of atoms or clusters of atoms
(small islands in semiconductor surfaces, molecules in classical
systems) at the surface.  
Second, for a practical reason: the
term surfactant has essentially become the accepted term
by virtue of its wide use in semiconductor growth, 
and the need for consistency with
existing literature forces it upon us.   
These reasons, we feel, justify and 
legitimize the use of the term surfactant in the present context.

Surfactants have been used to modify the growth mode of several
systems, including growth of metal layers in homoepitaxy
and heteroepitaxy.  In our view, the physics relevant to 
such systems is significantly different than in the case 
of semiconductors.  Supporting this view
is the fact that typically a small fraction of a monolayer is 
needed to produce the surfactant effect in metal growth, 
whereas in semiconductors typically a full monolayer of
the adsorbate species (or what is the equivalent 
to full substrate coverage, depending on the surface 
reconstruction) is required.  Presumably, in the case of 
metals, a small amount of the adsorbate is sufficient to 
induce the required changes in surface kinetics
\cite{fiorentini95,breeman96}, by altering
nucleation rates or step-edge barriers (for some representative
examples of metal-on-metal growth mediated by surfactants
see Table 1).  In semiconductors,
on the other hand, the entire surface must be covered
by the adsorbate in order for the required changes in 
energetics and kinetics to be obtained.
Due to this fundamental difference, in the present article 
we will concentrate on the surfactant effect in 
semiconductor systems. 

\sgs
\begin{table}
\label{Table1}
\begin{minipage}[t]{9cm}
\begin{tabular}{l|c|l} 
Film/Substrate  & Surfactant & Reference \\ \hline 
Ag/Ag(111) & Sb  &
\cite{vandervegt92,rosenfeld93,vrijmoeth94,meyer94,meyer95,%
meyer96}\\
\hline 
Cu/Cu(111) & O  & \cite{wulfhekel96} \\ \hline 
Fe/Au(111) & Au  & \cite{begley93} \\ \hline 
Fe/Cu(111) & C + O  & \cite{memmel95} \\ \hline 
Cu/Ru(0001) & O  & \cite{wolter93,schmidt93,schmidt94} \\ \hline 
Co/Cu(111) & Sb  & \cite{scheuch94} \\ 
           & Pb  & \cite{camarero94,camarero96} \\ \hline 
\end{tabular}
\caption{
Examples of metal-on-metal growth mediated by surfactants.       
}
\end{minipage}
\end{table}


In the following we first review the available information
on the subject, both from the experimental (section II) 
and the theoretical (section III)  
point of view. We then present some theoretical arguments that 
we have advanced in an effort to create a comprehensive picture 
of the phenomenon (section IV).  Finally, we discuss our views on 
remaining important issues for future research on 
surfactants and comment on 
prospects for their use in the fabrication of electronic
and optical semiconductor devices (section V).

\section*{II. Experimental Observations}

A wide range of systems have been studied where the surfactant
effect was demonstrated.  We classify these in three 
categories: the first consists of growth of group-IV layers
on group-IV substrates, the second of growth of III-V compounds 
on III-V substrates, and the third of mixed systems, including
growth of elemental and compound systems on various substrates.
This categorization has been inspired by the substrate features
and the nature of the deposited species, which together 
determine the growth processes.

\subsection*{1. Group-IV films on group-IV substrates}

In the first category, the substrate is either Si or Ge (in 
different crystallographic orientations), on which combinations
of different group-IV elements are deposited (Si, Ge, C). 
In these systems
the deposited species are mostly in the form of single group-IV atoms. 
The adsorbate layers consist of monovalent (H), trivalent
(Ga, In), tetravalent (Sn, Pb), pentavalent (As, Sb, Bi), 
and hexavalent (Te) elements, or noble metals (Au).
These adsorbates remove the usual reconstruction of the surface
(the different versions of the $(2 \times 1)$ reconstruction 
for the Si and Ge(100) surfaces,
the $(7 \times 7)$ and the c$(2\times8)$ for the Si and Ge(111)
surfaces, respectively),
and produce simpler reconstructions which are 
chemically passivated.  Characteristic examples are the 
$(1\times 1)$ reconstruction (induced by H or As on the (111)
surfaces), 
the $(\sqrt{3} \times \sqrt{3})$ reconstruction on the (111) surfaces 
with either one adsorbate atom per unit cell (induced 
by Ga or In) or three adsorbate atoms per unit cell
(induced by Sn, Sb, Pb, Au), the $(2\times1)$ reconstruction
of the (100) surfaces (induced by the trivalent and 
pentavalent elements or by H in the monohydride phase),
and the $(1\times 1)$ reconstruction of the (100) surfaces
(induced by Te or by H in the dihydride phase). In these
reconstructions
the dangling bonds of the substrate atoms are saturated
by the additional electrons of the adsorbate atoms,
producing low-energy, chemically unreactive surfaces.

One important issue in these systems is the strain  
induced by the deposition of atoms with different 
covalent radius than the substrate atoms.  The  
normal growth mode in strained systems involves the 
formation of 3D islands which relieve the strain
by relaxation at the island edges, either right from the 
initial stages of deposition (the so called Volmer-Weber
or 3D island growth), or after the formation of a wetting layer
(the Stranski-Krastanov growth). When surfactants are employed,
it is possible to induce layer-by-layer growth in strained 
systems by avoiding the formation of 3D islands for
film thicknesses much beyond what is obtained under normal 
conditions. The reduction of strain-induced islanding
was in fact one of the early intended
results of surfactant use, and remains a goal pursued
in several experimental studies. Even when surfactants
are used, however, and the 3D island mode of growth is
suppressed, the strain in the heteroepitaxial film is still 
present and is usually
relieved by the introduction of a network of misfit 
dislocations. The mechanism by which this happens is  
not known and remains to be analyzed by atomistic 
models.  

It is natural to expect that diffusion of  
group-IV adatoms on the adsorbate covered 
surfaces will be relatively easy due to the chemical 
passivation by the surfactant layer. Such a situation may lead to a
substantial increase of the diffusion length of adatoms on top of the
surfactant layer. Indeed, 
it has been reported experimentally that in Ge on Si 
heteroepitaxy certain elements, like Ga, In, Sn and Pb, 
lead to an increase in the width of the depleted zone around islands
\cite{voigtlander95}. At the same time, however, it has also been found
that other elements, like As, Sb, Bi and Te, lead to 
a decrease in the width of the depleted zone \cite{voigtlander95}. 
These observations were interpreted as indicative that the former
type of surfactant (group-III and group-IV atoms) enhance the 
diffusion length while the latter type of surfactants (group-V and
group-VI atoms) reduce the diffusion length. Moreover, this
interpretation has been frequently invoked as an explanation of the
suppression of 3D islanding in heteroepitaxy by group-V and group-VI 
surfactants.

Since it is generally easier for group-V and group-VI elements to 
provide a chemically passive surface, we argue that the above
interpretation may not be unique. In fact, we show in section IV
that even if it were true, it would not explain the surfactant effect
neither in homoepitaxy nor in heteroepitaxy. We propose an alternative
interpretation of the experimental results according to which the
diffusion length is mostly irrelevant. Instead, the essential question
is whether the surfactant layer passivates island edges or not.
Some surfactants (group-III and group-IV elements)
cannot passivate island edges which then act as strong sinks 
of newly deposited atoms, while other surfactants (group-V and 
group-VI elements) passivate island edges as well as  
terraces, so that island edges do not act as adatom sinks,
and the width of the depleted zone is reduced. We show that this
interpretation is consistent with the experimentally observed 
surface morphologies and island densities in 
the presence of surfactants. It also explains why group-V and 
group-VI adsorbates suppress 3D islanding in heteroepitaxy.
The systems studied experimentally
that belong to this category are listed in Table 2.  
The relative simplicity of the surface reconstruction induced 
by the surfactant and the fact that the deposited species is
mostly single group-IV atoms make the systems in this category
the easiest to analyze from a microscopic point of view.
Indeed, most atomistic scale models of the surfactant effect
address systems in this category.

\sgs 
\begin{table}
\label{Table2}
\begin{minipage}[t]{16cm}

\begin{tabular}{l|c|l} 
Film/Substrate  & Surfactant & Reference \\ \hline 
Si/Si(100)     &     H      &
\cite{copel94} \\ \hline 
Si/Si(111)     &    Ga      &
\cite{nakahara92,voigtlander93a,voigtlander94,hoegen95,voigtlander95,%
gallas97}\\
                &    In      & \cite{minoda93,voigtlander95} \\ 
                &    Sn      &
                \cite{iwanari90,iwanari91a,iwanari91b,iwanari92a}
                \\ 
                &    As      &
                \cite{voigtlander93a,voigtlander94,hoegen95,%
                voigtlander95} \\ 
                &    Sb      &
                \cite{voigtlander93a,voigtlander94,hoegen95,%
                voigtlander95} \\ 
                &    Au      & \cite{yagi92,wilk94}  \\ \hline  
Ge/Si(100)     &    H       &
\cite{sakai94,ohtani94,zaima95,ikarashi95,akazawa96,kahng97,kahng98}
\\ 
                &    B       & \cite{klatt94} \\ 
                &    In      & \cite{eaglesham93} \\ 
                &    Sn      & \cite{dondl92,lin95,lin96} \\ 
                &    As      &

                \cite{copel89,copel90,eaglesham91,copel91,tromp92,jusko%
                92,kohler91,ide95} \\
                &    Sb      &
                \cite{copel90,copel91,thornton91,thornton92,osten92a,os%
                ten92b,osten92c,osten92d,osten92e,cao92,krueger92,%
				sakamoto92}\\
		&            &
		\cite{eaglesham93,dondl92,%
		yang92,yang93,osten93,hoegen93b,hoegen94a,hoegen94b,%
                kissinger94a,kissinger94b,sasaki94}\\
		&      &
		\cite{boshart96a,boshart96b,katayama96,zhu97} \\ 
                &    Bi      &
                \cite{sakamoto93,sakamoto94,kawano93,katayama96} \\ 
                &    Te      &
                \cite{higuchi92,osten93,yang93,bennett97} \\ \hline
Ge/Si(111)     &    H       & \cite{terry93} \\ 
                &    Ga      &
                \cite{falta93,falta96,sakai94,voigtlander95,maruno96}
                \\ 
                &    In      & \cite{voigtlander95,minoda96} \\ 
                &    As      &
                \cite{yasuda95,voigtlander95,schellsorokin94,%
                               aizawa97} \\ 
                &    Sb      &
                \cite{hoegen91,legoues91,hoegen93a,hoegen93c,hoegen94b,%
                hoegen94c,hoegen94d,hoegen94e,larsson94,voigtlander94}\\
				&	&
                \cite{voigtlander95,voigtlander96,davoli97,reinking97}
                \\ 
                &    Bi      & \cite{minoda97} \\ 
                &    Au      & \cite{yagi92} \\ \hline  
Si/Ge(100)     &    H       & \cite{akazawa97} \\ 
                &    Sb      & \cite{yang93,akazawa96} \\ \hline  
Si/Ge(111)     &    Ge      & \cite{lin94} \\ \hline 
Si$_{1-x}$C$_x$/Si(100)    &    Sb      &
\cite{pettersson95,pettersson97} \\ \hline 
Ge$_{1-x}$Si$_{x}$/Si(100)   &    H       & \cite{chelly97,wado95} \\ 
                &    Sn      & \cite{lin95,wakahara95} \\ 
                &    As      & \cite{xie97} \\ 
                &    Sb      & \cite{li95,nilsson96} \\ \hline  
Si/SiGe(100)   &    As      & \cite{copel90}  \\
                &    Sb      & \cite{copel90}  \\ \hline 
Ge$_{1-x}$C$_x$/Si(100)    &    Sb      & \cite{osten94} \\ \hline  
Si/Si$_{1-x}$Ge$_x$(100)    &    H      & \cite{ohtani93} \\ \hline  
Sn/Ge(100)     &    Sb      & \cite{dondl97} \\ \hline  
Si$_m$Ge$_n$/Si(100)   &    Sb      & \cite{presting92} \\ \hline  
Si$_{1-x-y}$Ge$_{x}$C$_{y}$/Si(100)  &    Sb      & \cite{croke97} \\
\hline  
\end{tabular}
\caption{
Systems in the first category of surfactant mediated semiconductor
growth:
group-IV films on group-IV substrates.
}
\end{minipage}
\end{table}

\subsection*{2. III-V films on III-V substrates}
     
The second category consists of III-V substrates on 
which combinations of other III-V systems are deposited.
The deposited species in this case are more complicated,
since at least two types of atoms have to be supplied 
with different chemical identities.  Under usual conditions
the group-III species is deposited as single atoms,
whereas the group-V species is deposited as molecules (dimers 
or tetramers), which have to react with the group-III atoms
and become incorporated in the growing film.  This is 
already a significant complication in growth dynamics,
and makes the construction of detailed atomistic growth models
considerably more difficult.  Moreover, the usual surface 
reconstructions of these substrates are more complicated 
and depend on deposition conditions (temperature
and relative flux of group-III to group-V atoms).  In the presence of
surfactants both the surface reconstructions and the 
atomic motion is altered, but much less is known about 
the atomic level details.  The surfactants used in these systems
include H, Be, B, In, Sn, Pb, As, Sb, Te.  In certain cases,
the surfactant species is the same as one of the atoms in the 
growing film (such as In in InAs growth on GaAs), or one of the atoms
in the substrate (such as Sb in growth of InAs on AlSb).  
Strain effects are important in these systems as well.
Building high quality III-V heterostructures has been one of the 
goals of many technologically oriented studies, and the
use of surfactants has been beneficial in reducing the
problems associated with strain.  
However, the more 
complex nature of these systems has prevented 
detailed analysis of the type afforded in group-IV systems.
A compilation of experimental results for this category
is given in Table 3. 

\sgs 
\begin{table}
\label{Table3}
\begin{minipage}[t]{10.5cm}
\begin{tabular}{l|c|l} 
Film/Substrate  & Surfactant & Reference \\ \hline 
GaAs/GaAs(100)  &    H       &
\cite{okada95a,okada95b,kawabe95,okada96} \\ 
                &    Sn      & \cite{massies93} \\ 
                &    Pb      & \cite{massies93} \\ 
                &    Te      & \cite{massies93,grandjean96} \\ \hline 
InAs/GaAs(100)  &    H       & \cite{yong95} \\ 
                &    In      & \cite{behrend96} \\ \hline 
AlGaAs/AlGaAs(100)&  Be      & \cite{cunningham95} \\ 
                  &  Sb      & \cite{kaspi95a,kaspi95b} \\ \hline  
InGaAs/GaAs(100)&    Sn      & \cite{petrich91} \\ 
                &    Te      &
                \cite{massies92,grandjean92,grandjean93,snyder93}\\
                \hline
GaAs/InP(100)   &    H       & \cite{yong95} \\ \hline  
InAs/InP(100)   &    H       & \cite{chun96} \\ \hline  
InGaAs/InP(100) &    H       & \cite{lapierre97} \\ \hline
GaAs/GaAs(111)  &    In      & \cite{ilg93} \\ \hline  
GaAs/GaInAs(100)&    Te      & \cite{delamarre97} \\ \hline  
InAs/GaInAs(100)&    In      & \cite{tournie94a} \\ \hline  
GaN/GaN(100)    &    As      & \cite{feuillet97} \\ \hline  
GaN/AlGaN(100)  &Si(CH$_3$)$_4$ & \cite{tanaka96,tanaka97} \\ \hline 
InAs/AlInAs(100)&    In      & \cite{tournie93a,tournie93b} \\ \hline  

InAs/AlSb(100)  &    Sb      & \cite{tuemmler97} \\ \hline  

InAs/InPSb(100) &    Sb      & \cite{tuemmler97} \\ \hline  
InAs/InGaAs(100)&    In      & \cite{tournie92,tournie95,ploog94} \\
\hline  
\end{tabular}
\caption{
Systems in the second category of surfactant mediated semiconductor
growth:
III-V films on III-V substrates.
}
\end{minipage}
\end{table}

\subsection*{3. Mixed film and substrate systems}

The final category consists of mixed systems in which group-IV 
films are grown on III-V substrates (for example Si on GaAs)
or vice versa (for example GaN on Si).  In these systems,
in addition to the usual strain effects one has to consider 
also polarity effects, which arise from the fact that at the
interface different types of atoms are brought together
and their dangling bonds contain different amounts of 
electronic charge which do not add up to the proper value 
for the formation of covalent bonds.  It is possible that 
the surfactant layer plays an important role in reducing 
polarity problems, as well as modifying the energetics
and suppressing strain effects, as it does in the previous 
two categories.  

Since the substrate and the thin film are rather different 
for systems in this category, we include in the same 
category a number of odd systems which involve the 
presence of insulating buffer layers (like CaF$_2$ in 
the growth of Ge on Si substrates) and the growth of metal
layers (like In on 
Si, In on GaAs, Sn on Ge and Ag on Si) 
or silicide layers (like CoSi$_2$ on Si),
as well as the growth of technologically important semiconductors
on insulators (like GaN on sapphire).  All these cases 
are important for device applications and it is interesting
to study how surfactants can be employed to improve the 
quality of growth.  However, the complexity of the structures
involved and the several different species of atoms present
make the detailed analysis of systems in this category rather
difficult. We tabulate the experimental information 
for systems in this category in Table 4. 

\sgs 
\begin{table}
\label{Table4}
\begin{minipage}[t]{9cm}
\begin{tabular}{l|c|l} 
Film/Substrate  & Surfactant & Reference \\ \hline 
Si/GaAs(111),(311) &  H  & \cite{peng97,zhang97} \\ \hline 
Si/GaAs(100)      &  As  &  \cite{avery95,sudijono96} \\ \hline
Ge/InP(100)      &  Sb  &  \cite{rioux91,rioux92} \\ \hline
Ge/GaF$_2$/Si(100),(111) &   B  &  \cite{cho93} \\ \hline
GaAs/Ge(100)   &    H    & \cite{okada97} \\ \hline
CoSi$_2$/Si(100) &  As   & \cite{scheuch97} \\ \hline
Ag/Si(111) &  Sb   & \cite{park97} \\ \hline
In/Si(111)   &   H   & \cite{leisenberger97} \\ \hline 
In/GaAs(100) &   Sb  &  \cite{schintke97} \\ \hline 
Fe/Ge(100) &  S   & \cite{anderson96} \\ \hline
MnSb/GaAs(100) &  H   & \cite{ikekame97} \\ \hline
GaN/Si(111)    &   As    & \cite{mendoza97} \\ \hline
GaN/SiC(100) &  As   &  \cite{fueillet97} \\ \hline 
GaN/Al$_2$O$_3$& Bi  &  \cite{klockenbrink97} \\ \hline 
\end{tabular}
\caption{
Systems in the third category of surfactant mediated semiconductor
growth:
mixed films and substrates.
}
\end{minipage}
\end{table}

\section*{III. Theoretical Models}

Following the experimental 
observations, a number of theoretical models have been studied 
in order to understand and explain the surfactant effect in 
semiconductor growth.  We divide 
these models in three categories.  In the first category we place
models
that have concentrated on the microscopic aspects, attempting 
to understand the atomic scale features and processes involved
in this phenomenon; models of this type typically employ sophisticated
quantum-mechanical calculations of the total energy in order to 
evaluate the relative importance of the various structures, and 
in order to determine the relevant activation energies involved in 
the kinetic processes.  In the second category we place models
that are more concerned with the macroscopic aspects of the 
surfactant effect, such as island morphologies and distributions
as well as the effects of strain, without attempting to explain 
the details of the atomistic processes, although these may be taken 
into account in a heuristic manner.  Finally, in the third category
we place models that attempt to combine both aspects, that is, 
they try to use realistic descriptions of the atomistic processes
as the basis for macroscopic models.  Evidently, this last type 
of models is the most desirable, but also the most difficult to 
construct. We review models in these three categories in turn.

\subsection*{1. Microscopic models}

Initial attempts at understanding the microscopic aspects of
surfactant mediated growth focused on the thermodynamic aspects, 
that is, strived to justify why it is reasonable to expect 
the surfactant layer to float on top of the growing film.
This was investigated by calculating energy differences between 
configurations with the surfactant layer buried below layers of 
the newly deposited atoms versus configurations with the 
surfactant on top of the newly deposited atoms \cite{copel89}.
These energetic comparisons, based on first-principles calculations
employing density functional theory, established that there exists
a strong thermodynamic incentive for keeping the surfactant layer
on top of the growing film.

In a similar vein, calculations by Kaxiras \cite{kaxiras93}
established that certain surfactants are more likely to lead to 
layer-by-layer growth than others, while a simplistic analysis 
of their chemical nature would not reveal such differences. These
calculations were done using different types of surfactants
on the same substrate and considering the relative energies of the 
various surface reconstructions induced by the surfactant layer. 
Specifically, three different types of group-V atoms were considered
as surfactants, P, As and Sb, on the Si(111) surface for Ge or Si
growth.  The similar
chemical nature of the three elements would argue for very similar
surfactant behavior.  However, the total energy 
calculations indicated that the three elements give significantly 
different reconstructions, some of which would lead to relatively
easy floating of the surfactant layer, while others would hamper
this process.  This is related to the manner in which, in a given 
reconstruction, the surfactant atoms are bonded to the substrate.
For instance, in the energetically preferred reconstructions,
the P and As atoms are bonded to the substrate with three strong 
covalent bonds each, while Sb atoms are bonded to the substrate 
with only one strong covalent bond per adsorbate atom.  Based on
these comparisons, Kaxiras proposed that Sb would work well on 
this substrate as a surfactant, while P and As would not 
\cite{kaxiras93,kaxiras95}, 
a fact that was subsequently verified experimentally \cite{hoegen95}.  

A study by Nakamura et al. \cite{nakamura96} of the same system
(the Si(111) substrate with Sb as surfactant for Ge growth), 
based on the discrete variational
approach and the cluster method to model the surface, 
reported  
that the presence of the surfactant strengthens the bonds between 
the Ge atoms on the surface.  This effect, it was argued, leads to 
nucleation of stress-relieving dislocations at the surfaces which 
is beneficial for layered growth of defect-free films.  In this analysis
neither the defects themselves
nor any type of exchange and nucleation mechanisms was considered
explicitly.  Moreover, the bond-strengthening arguments are of
a chemical nature, which may be useful in a local description of
chemical stability, but sheds little light on the dynamics 
of atoms during surfactant mediated growth.
The chemical nature of Sb bonding on the Si(100) and 
the Ge(100) substrates was 
also investigated by Jenkins and Srivastava \cite{jenkins96,jenkins97}.
In this work, first-principles density functional 
theory calculations were employed to determine the structure and
the nature of bonding of Sb dimers in the $(2 \times 1)$
reconstruction,
which though interesting in itself, provides little  
direct insight into the process of surfactant mediated growth. 

The theoretical models considered so far addressed the problem 
of surfactant mediated growth by considering what happens at the 
microscopic level, but for entire monolayers, that is by imposing
the periodicity of the reconstructed surface in the presence of
the surfactant.  Subsequent atomistic models studied
the equilibrium configurations and 
dynamics of individual adsorbed atoms or dimers, which 
is more appropriate for understanding the nature of growth on the 
surfactant-covered surface.  A first example was an attempt 
by Yu et al.\
\cite{yu94a,yu94b,yu96} to justify how 
newly deposited Ge atoms on the As-covered Si(100) surface
exchange place with the surfactant atoms in order to become
embedded below the surfactant layer. 
In these calculations, based on total energy comparisons obtained
from density functional theory, the metastable and stable 
positions of Ge dimers are established, indicating possible 
paths trough which the newly deposited Ge atoms can be 
incorporated under the surfactant As layer.  However, no 
explicit pathways were determined, and therefore no 
activation energies that might be relevant to growth 
kinetics were established.  Furthermore, even though specific 
mechanisms for growth of needle-like islands by appending Ge 
dimers to a seed are discussed, the lack of calculated energy barriers
for the exchange process 
and a large number of unproven assumptions involved in
the proposed mechanisms, means they are of little help in 
understanding the surfactant effect.  For instance, the 
work of Yu et al.\ \cite{yu94a} claims that Ge dimers are actually 
situated between As dimer rows instead of on top of the As dimer rows, 
while in their proposed island growth mechanism they employ
configurations
that involve Ge dimers on top of the As dimer rows.   

A similar type of analysis by Ohno\ 
\cite{ohno94,ohno96},  
also using density functional theory
calculations of the total energy, was reported for Si-on-Si(100)  
homoepitaxy using As as surfactant. 
Ways of incorporating the
newly deposited Si dimers below the As layer were considered
by studying stable and metastable positions, and the rebonding 
that follows the exchange process.  Again, though, actual exchange 
pathways and the corresponding activation energies relevant to 
growth kinetics, were not considered.  
In this study it is shown explicitly how exchange of isolated
Si dimers on top of the As layer is not exothermic, while the 
presence of two Si dimers leads to an energetically preferred
configuration after exchange. 
This fact is used to argue that the Si dimer interactions are
responsible for both their mutual repulsion and the initiation 
of the exchange.  It appears however that these two effects,
that is, the strong repulsion of ad-dimers and the requirement of
their presence at neighboring sites for the initiation of exchange,
would be incompatible as far as growth is concerned.
Both the work of Yu et al.\ 
\cite{yu94a,yu94b,yu96} and the work of Ohno \cite{ohno94,ohno96}
deal with mechanisms in which the basic unit involved in the exchange  
process is a deposited dimer,
as was originally suggested by Tromp and Reuter \cite{tromp92}.    

An interesting microscopic study of surfactant mechanisms was 
reported by Kim et al.\ \cite{kim96,oh96,kim97}.  In this work,
first-principles
molecular dynamics simulations were employed to investigate the 
effect of Sb atoms at step edges on the Si(100) surface for Si 
homoepitaxy.  
This study examined the effect of Sb dimers
on the step-edge barriers (also referred to as Schwoebel-Ehrlich 
barriers \cite{schwoebel,ehrlich}, 
for which we adopt here the acronym SEB which is both 
descriptive and referential).  
These are extra barriers to adatom attachment to the step-edge
when the adatom arrives from the upper terrace,
compared to the barriers for diffusion on the flat terraces.
The authors find that the presence of Sb at the step edge
gives a significant SEB for the attachment of a single Si atom,
but much smaller SEB for attachment of a Si dimer by the 
push-over mechanism (in which the Si dimer at the upper terrace pushes
the Sb 
dimer at the step edge over by one lattice constant, and thus becomes 
incorporated in the bulk).
This relative suppression of the SEB for dimer attachment 
leads to layer-by-layer growth as opposed to 
3D island growth, and consequently, Kim et al.\ argue, the presence of 
the surfactant Sb dimer at the step edge
would lead to layered growth. This is an interesting
suggestion, but it remains to be proven that it is the 
correct view for the system under consideration.  Specifically,
it is not clear whether a configuration with Sb dimers only
at step edges of the Si(100) surface is stable.  
Typically, an entire monolayer is needed
for the surfactant effect in similar systems, and the precise coverage
is a crucial aspect of the effect.  If the surfactant coverage 
is different than that assumed in the model of Kim et al.\ 
then the atomic processes at the step edge could be very different
leading to a different picture of the effect.
Furthermore, kinetic Monte Carlo studies are required to establish
that the calculated energy barriers can actually lead to the
predicted mode of growth, since the density of Si adatoms (determined
by the flux, the diffusion rate and the attachment-detachment
rates) will also influence the growth process.  

Another detailed study of activation energies for diffusion and
exchange
processes in surfactant mediated epitaxy was reported by Ko 
et al.\ \cite{ko96}. This study was also based on first-principles
calculations
of the energetics and addressed Si epitaxy on Si(100) with As acting 
as surfactant. In this work it was established that the
exchange of a Si adatom with a sublayer As site involves an 
energy barrier of 0.1 eV, which is considerably lower than 
the energy barrier for diffusion (of order 0.5 eV) or the
energy barrier for dimer exchange (of order 1.0 eV) which had been
invoked as a possible mechanism in earlier studies of the same system
\cite{tromp92,yu94a,yu94b,yu96,ohno94,ohno96}. 
This is a very interesting suggestion, but falls short of  
providing a complete picture of the surfactant effect.
Specifically, it is not clear how a single exchange step of
the type investigated in the study of Ko et al.\ \cite{ko96}
can lead to a configuration that will nucleate the next 
layer of the crystal. This process may well involve additional 
important steps with different activation barriers, so that the 
barrier calculated, though important and interesting, may not 
be the determining step in the growth process.  
In fact, Ko et al.\ find that exchange of two individual Si atoms
at neighboring sites leads to the formation of a protruding 
As dimer, which acts as a seed for further growth.
This protruding As dimer binds additional Si adatoms and 
leads to the formation of a Si dimer, which eventually undergoes
site exchange with a neighboring As dimer with an energy barrier
of 1.1 eV.  It would then appear that it is this last step
that is the determining step in growth, which leads back to 
the dimer-exchange picture discussed earlier, albeit now with  
a more detailed picture of how this process may be initiated
by the barrierless exchange of single Si adatoms. 

Two separate studies of growth on III-V surfaces addressed the 
surfactant effect in these systems. 
In the first study, by Miwa et al.\ \cite{miwa96},  
the dimer exchange mechanism on III-V surfaces, using Te as the 
surfactant, was investigated using first-principles calculations. 
These authors find that InAs growth 
on GaAs(100) proceeds by complete dimer exchange between the In and
Te layers on the As-terminated surface, while on the In-terminated 
surface the exchange between the Te layer and an overlayer of As 
is only partial.  
The second study, by Shiraishi and Ito \cite{shiraishi96,shiraishi98}, 
examined the equilibrium configurations of adatoms 
on the GaAs(100) surface with different 
As coverages, using first-principles total energy calculations. 
This study concluded that preadsorbed Ga atoms
play a ``self-surfactant'' effect by significantly influencing 
the adsorption energy of As dimers at various sites on the surface.
In this analysis, energy barriers for diffusion and exchange mechanisms
are not taken into account, and consequently the interpretation of 
actual growth processes is limited.  

The most detailed study of actual atomistic mechanisms for diffusion 
and exchange was reported by Schroeder et al.\ \cite{schroeder98}.
This work examined the motion of Si adatoms on the As covered Si(111)
surface, using first-principles total-energy calculations.  The authors
report a very interesting pathway for exchange between the additional
Si
atom and an As surfactant atom with an energy barrier of only 0.27 eV.
This is comparable to the diffusion barrier for the Si atom on top of
the As layer, calculated to be 0.25 eV.  The Si atom can 
undergo the reverse of the exchange process, 
and by so doing it can 
get on top of the As surfactant layer, 
a process that involves an energy barrier
of 1.1 eV, according to the results of Schroeder et al.  
This leads to a rather complex sequence of events,
with Si adatoms arriving at
the surfactant covered substrate, diffusing, exchanging, undergoing the 
reverse process 
and diffusing again, with the relevant energy barriers.  
The possibility of 
the reverse of the exchange process was first explicitly introduced 
in the work of Kandel and Kaxiras \cite{kandel95}, 
where it was called ``de-exchange''.
We adopt this term in the following 
as more descriptive of the reverse of the 
exchange process, since this undoes the effect of an exchange
step rather than repeat it, as the term   
``re-exchange'' (used 
in the work of Schroeder et al.) might suggest.
The de-exchange process was shown by Kandel and Kaxiras to
be a crucial process in maintaining the layer-by-layer growth mode in 
the presence of the surfactant (see more details below).  
This de-exchange process had been found to have a higher 
activation energy (1.6 eV) than either exchange (0.8 eV) or diffusion
(0.5 eV) by Kandel and Kaxiras, 
although this was established by considering the exchange
or de-exchange of entire monolayers of newly deposited atoms on
top of the surfactant layer. Schroeder et al.\ show that the same
energy 
ordering is valid 
also for individual adatoms on top of the surfactant layer, but the 
actual barriers for individual adatoms are lower (1.1 eV, 0.27 eV and 
0.25 eV for de-exchange, exchange and diffusion, respectively).
This establishes unequivocally the importance of the de-exchange
process.  
What is lacking from the work of Schroeder et al.\ is a sequence of
steps that can actually lead to the formation of the next layer of
deposited material.  Specifically, even after the single Si adatom 
has exchanged positions with a surfactant As atom, the system is not
in a configuration from which the repeated sequence of similar steps 
could lead to the formation of a new layer.  In the system studied by
Schroeder et al.\ this process may be quite complicated, since the 
Si(111) surface consists of double layers, the formation of which 
may involve additional energy barriers which supersede the one 
determined for the exchange of a single adatom. 

\subsection*{2. Macroscopic models}

There is a debate in the literature on whether the suppression of 3D
islanding by surfactants in heteroepitaxy is an equilibrium effect or a
kinetic one. While most researchers in the field take the kinetic
approach, there has been some effort to try and explain the surfactant
effect using thermodynamic considerations. According to the
thermodynamic
approach, the equilibrium state of the newly deposited material in the
presence of a surfactant layer is a smooth flat film. The underlying 
assumption behind kinetic models is that even with surfactants, the
true
equilibrium state of the system is that of 3D islands. The role of
surfactants, in this case, is to induce layer-by-layer growth
kinetically and to make the approach to equilibrium longer than
realistic time scales. We will first give examples of the thermodynamic
approach to the surfactant effect and then elaborate on some kinetic
models.

Kern and M\"{u}ller \cite{kern95} calculated the free energy
of formation of a crystal of material A stretched to be coherent with a
substrate of material B. They took into account effects of surface
energy as well as surface stress and obtained the equilibrium shape of
the crystal by minimizing its free energy with respect to its height
and width. In their view, surfactants may reduce surface stress and
surface energy, and hence lead to flatter islands and maybe even to
wetting of the substrate by the deposited material (which happens when
the equilibrium island height vanishes). They view such surfactant
induced wetting as a transition from 3D growth to 2D layer-by-layer
growth. This Kern-M\"{u}ller criterion may serve as an indication of
whether the effect of a certain surfactant is in the right direction to
suppress 3D islanding. However, they do not consider the possibility of
3D growth when the deposited material wets the substrate
(Stranski-Krastanov growth mode). They also ignore strain relaxation,
which reduces the cost of 3D island formation. Thus, the question of
whether the surfactant effect could be a purely thermodynamic one is
left unanswered.

A different equilibrium argument was proposed by Eaglesham et al.\
\cite{eaglesham93}. These authors argue that surfactants change the
surface
energy {\em anisotropy} and this leads to the suppression of 3D
islanding. They examine experimentally islands of Ge on Si(100) films
with 
and without surfactants, and find that their equilibrium shape changes
radically in the presence of surfactants and depends strongly of the
specific surfactant used. For example, Sb as a surfactant favors (100)
facets, whereas In favors (311) facets. They advance the idea that if
the surfactant favors facets in the same orientation as the substrate,
the equilibrium shape of the islands generated will be flat. This will
lead to earlier coalescence of islands and will enhance layer-by-layer
growth. The mechanism proposed by Eaglesham et al.\ may have a
significant impact on the growth mode. But it cannot be the main
explanation of the surfactant effect, since the equilibrium
morphologies
observed in their experiments include 3D islands. Therefore, in their 
explanation of the surfactant effect they supplement the equilibrium
consideration with a kinetic one, i.e. the reduction of the diffusion
length induced by surfactants.

It seems quite difficult to explain the surfactant effect relying on
thermodynamics alone. For this reason most researchers in the field
make the assumption that surfactants suppress 3D islanding {\em
kinetically}. Markov's work \cite{markov94,markov96,markov97} 
is an example of such a
kinetic model. He developed an atomistic theory of nucleation in the
presence of surfactants. The main results of this work
are expressions for the
nucleation rate and saturation density of islands. These quantities
depend crucially on the difference between the energy barrier for
adatom
diffusion on top of the surfactant layer and the barrier for diffusion
on a clean surface. If this difference is positive, 
surfactants decrease the diffusion length for adatoms and the
saturation
density of islands rises sharply. Such an anomalously high island
density
in the presence of surfactants has been seen experimentally in
various systems, and is viewed by many researchers
\cite{voigtlander95,voigtlander93a,voigtlander94,%
eaglesham91,tromp92,osten93,%
grandjean96,massies93,ko96,sakai93}  
as the main mechanism by
which surfactants change the growth mode of the film and suppress 3D
islanding in heteroepitaxy. We will show in section IV that
this mechanism does not explain the surfactant effect.

An entirely different approach was taken by Barabasi
\cite{barabasi93a,%
barabasi93b}. Rather than looking at the kinetics of the system on the
atomic length scale, he viewed the growing film on a much coarser
scale. He
represented the local height of the film and the local width of the
surfactant layers as continuous fluctuating fields, in the spirit of the
KPZ model of kinetic roughening \cite{kpz}. Based on the
relevant symmetries of the system he wrote down a set of coupled 
differential equations which describe the dynamics of these two fields.
The quantity of interest in this approach is the width of the film
surface and its dependence on system size. Typically, such a theory
would predict a rough surface where the width diverges with system
size.
Barabasi found that surfactants can induce a flat phase where the
surface width does not diverge with system size. He associated this
phase with a layer-by-layer growth mode. It is interesting to see that
a theory on such a macroscopic length scale can capture effects which
depend
critically on processes that occur on an atomic scale. The drawback of
this
theory is that it is not clear what role the lattice mismatch and
strain 
play in the kinetics of the system. Also, in the rough phase the
model predicts a self-similar structure for the surface.
Experimentally,
however, the morphology of a surface with 3D islands is not self-%
similar, and it is not clear whether this continuum theory can describe
the experimental morphologies.
Barabasi and Kaxiras \cite{barabasi96} extended this model to include
two different dynamical fields, one representing the surfactant 
layer, the other
the surface film layer.  This allowed an investigation of whether
subsurface diffusion, which had been neglected in the previous model,
could change the behavior.  It was found that subsurface diffusion 
essentially always leads to roughening, and if it were operative in 
real systems it would prevent layer-by-layer growth.  

Most models of surfactant mediated epitaxial growth emphasize the
significance of adatom diffusion for the determination of the growth
mode.
Another atomic process of importance is attachment and detachment of
adatoms from island edges. In fact, in section IV we develop
a model according to which surfactants suppress 3D islanding by
passivating island edges, thus suppressing adatom detachment. It is
therefore of interest to investigate the influence of island-edge
passivation on surface morphology. 
Kaxiras first introduced the idea of island-edge passivation by the 
surfactant \cite{kaxiras96a,kaxiras97}, and carried out kinetic 
Monte Carlo simulations on a very simple model to show that it
can lead to morphologies compatible with experimental observations.  
Kandel also carried out a study of island-edge passivation effects
\cite{kandel97}. He investigated a simple model of submonolayer
homoepitaxial
growth in the framework of rate equation theory using the critical
island 
approximation (only islands of more than $i^\ast$ atoms are stable,
while
smaller islands decay). The main result of this work is that the island
density scales with flux, $F$, as $F^\chi$ with
$\chi=2i^\ast/(i^\ast+3)$ when island edges are passivated, while 
$\chi=i^\ast/(i^\ast+2)$ without island-edge passivation. This
conclusion is important because the exponent $\chi$ can be measured
experimentally and one can learn from its value whether island-edge
passivation is operative or not 
in the experimental system at hand. For example, a value of
$\chi>1$ can occur only if the surfactants passivate island edges. 
Kandel's theory relies on a somewhat oversimplified
picture of submonolayer growth, and the conclusions are yet to be
verified with a more rigorous theory or by detailed simulations of the
growth process.

\section*{IV. The Diffusion--De-exchange--Passivation Model}

To our knowledge the only attempt to construct a comprehensive 
model that includes both the microscopic aspects of atomic motion 
and a realistic description of the large length-scale evolution of
the surface morphology has been reported by the present authors
\cite{kandel95}. 
The work of Zhang and Lagally \cite{zhang94} is another attempt to link
the microscopics and the macroscopics of the effect of surfactants
on thin film growth. However,
their work discusses homoepitaxial growth of metals, a subject which is
interesting in its own right, but is beyond the scope of the present
review
article.

\subsection*{1. General considerations}

Before we embark on the construction of the theoretical model
\cite{kandel95},
we briefly review the relevant experimental information since 
we do not claim that a single model captures every type of 
surfactant mediated growth mode. We focus here on growth of
elemental or compound semiconductors, in which a single species 
of atoms controls the diffusion and exchange (or de-exchange)
processes, and the surfactant produces a chemically passivated surface.
We take as the canonical case a group-IV substrate (examples are
Si(100) or 
Si(111)) and a group-III or group-V surfactant (these are actually
the systems that have been studied most extensively experimentally,
as is evident from Tables 2-4).

It appears that a full monolayer of surfactant coverage is required for
growth of high quality semiconductor crystals.
This is different from the case of surfactant effects in the 
growth of metals, where a small amount of surfactant (typically
few percent of a monolayer coverage) is sufficient. 
The most direct evidence on this issue was provided by the
experiments of Wilk et al.\ \cite{wilk94}, who studied 
homoepitaxial growth of Si on Si(111) using Au as a surfactant.
These authors report that the density of defects in the film correlates 
well with the surfactant coverage, with the minimum defect 
density corresponding to full monolayer coverage 
by the surfactant. This is a physically
appealing result, and can be interpreted as evidence that 
the better the passivation of the surface by the surfactant,
the more effective the surfactant is in promoting high quality growth.
In the following we will assume that full monolayer coverage of
the substrate is the standard condition for successful 
surfactant mediated growth of semiconductors.

The model we will now describe assumes the surfactant effect is kinetic
in nature. As with all other kinetic models of surfactant mediated
growth, the underlying idea is that at equilibrium the 
heteroepitaxial system generates 3D islands even in the presence of
surfactants. The role of surfactants is to make the approach to
equilibrium very slow, so that 3D islands are not generated during the
growth of the film. This means that surfactants kinetically 
suppress one or more microscopic
processes, which are essential for the growth of 3D islands. The most 
important ingredient of any explanation of the surfactant effect is the 
identification of these processes. Almost all the explanations found
in the literature identify the relevant process as adatom diffusion 
\cite{voigtlander95,voigtlander93a,voigtlander94,%
eaglesham91,tromp92,osten93,%
grandjean96,massies93,ko96,sakai93}. 
The idea is that the
energy barrier for exchange of an adatom with a surfactant atom,
$E_{ex}$, is
smaller than the barrier for diffusion of an adatom on top of the
surfactant layer, $E_{d}$. 
An adatom therefore diffuses a very short distance
before it exchanges, and after exchange it cannot diffuse (once 
it is underneath
the surfactant layer). This suppressed diffusion mechanism explains the
surfactant effect in the following way: the reduced diffusion length
makes the density of islands nucleating on
the surface very high. As a result, island coalescence occurs before
any second layer islands nucleate on top of existing first layer
islands. This is how, according to this mechanism,
3D islanding is suppressed.

As mentioned in section II, the support for this
hypothesis comes from various experiments and particularly those of 
Voigtl\"{a}nder et al.\ \cite{voigtlander95}. In these experiments, they
studied the
effects of various surfactants on submonolayer homoepitaxial growth of
Si on Si(111). The results were correlated with studies of the effect
of the same surfactants on heteroepitaxial growth of Ge on Si(111).
Voigtl\"{a}nder et al.\ found that generally there are two types of
surfactants. Group-III and group-IV elements tend to significantly
decrease the island density in submonolayer homoepitaxy and lead to 3D
islanding in heteroepitaxy. On the other hand, group-V and group-VI 
elements drastically increase the island density in submonolayer
homoepitaxy and suppress 3D islanding in heteroepitaxy. If one
interprets an increase in the island density as an indication of
suppression of diffusion, these results confirm the mechanism discussed 
above.

Despite the appealing nature of the suppressed diffusion hypothesis, we
have proposed that it may not be the entire story. 
Our concerns arose from the fact that 
group-V and group-VI elements chemically passivate the surface more
efficiently than group-III and group-IV elements. Intuitively, this
should lead to faster diffusion on surfaces covered by the former
elements. But the experimental results are consistent with the latter
elements enhancing diffusion and the former ones suppressing it. To
clarify this issue, we decided to examine more carefully the microscopic
processes involved in the kinetics of surfactant mediated epitaxy. Our
investigation led to an entirely different explanation of the influence
of surfactants on epitaxial growth modes.

A schematic representation of the possible atomic
processes is shown in Fig.\ 1. The simplest process is of course 
diffusion of adatoms on top of the surfactant 
layer [Fig.\ 1(a)]. A second important process is the exchange of
adsorbed atoms with the surfactant atoms, so that the 
former can be buried under the surfactant layer and become part
of the bulk. This process can take place either on a terrace
or at a step [Fig.\ 1(b)]. From thermodynamic considerations, 
we must also consider the process by which atoms de-exchange and 
become adatoms which can diffuse on top of 
the surfactant layer [Fig.\ 1(c)]. Again, this process can 
take place on terraces or at surface steps.
Finally, we have to consider separately the case 
of surfactants that {\em cannot} passivate step edges,
in which case both the exchange [Fig.\ 1(d)] 
and de-exchange processes [Fig.\ 1(e)] will be different than 
at passivated steps, since they no longer involve actual 
exchange events between adatoms and surfactant atoms. 
We refer to our model as the Diffusion--De-Exchange--Passivation (DDP)
model, since these are the three processes that determine the 
behavior in surfactant mediated epitaxy:
diffusion is always present; de-exchange obviously implies also the 
presence of exchange; and passivation (always present on terraces)  
may or may not be present at island edges, 
but either its presence or its absence
is a crucial element.

\subsection*{2. First-principles calculations}

In order to evaluate the relative contributions of these processes and
their influence on the growth mode, the 
corresponding activation energies must be calculated.
This is a difficult task because very little is known 
about the atomic configurations involved.
We therefore begin by considering two idealized processes that
involve entire monolayers, 
discuss how the corresponding activation energies could
be representative and relevant for growth mechanisms, and obtain their
values from first-principles calculations.

The first process we consider is diffusion on a surface
covered by a surfactant monolayer. The representative system
we chose to study consists of a Si(111) substrate, 
covered by a bilayer of Ge, with Sb as the surfactant.
In this case, it is known that the structure of the Sb 
layer is a chain geometry with a periodicity of $(2\times1)$ 
as shown in Fig.\ 2 \cite{kaxiras93}. An additional Ge atom is then 
placed on top of the Sb layer and the energy is optimized
for a fixed position of the Ge atom along the direction
parallel to the Sb chains. All other atomic coordinates,
including those of the Ge atom perpendicular to the Sb chain
and vertical with respect to the surface, were allowed to relax
in order to obtain the minimum energy configuration.
The energy and forces were computed in the framework of 
Density Functional Theory and the Local Density Approximation 
(DFT/LDA), a methodology that is known to provide accurate
energetic comparisons for this type of system
(see in particular the reviews by Kaxiras \cite{kaxiras96b,kaxiras96c}
on the application of 
such calculations to semiconductor growth phenomena).
By considering several positions of the extra Ge atom
along the chain direction and calculating the corresponding 
total energy of the system, we obtained a measure of the 
activation energy for diffusion in this direction.
We found that the activation energy for diffusion along this path is 0.5
eV.

We next considered a possible exchange mechanism in the same system, 
through which the newly deposited Ge atoms can interchange 
positions with the surfactant atoms and become buried under them.
To this end, we modeled the system by a full monolayer of Ge
deposited on top of the surfactant layer [Fig.\ 3(a)].  We studied 
a concerted exchange type of motion for the Ge-Sb interchange.
In the final configuration [Fig.\ 3(e)] the Ge layer is below
the Sb layer, and the system is now ready for the deposition of
the next Ge layer on top of the surfactant. The middle 
configuration, Fig.\ 3(c), corresponds to a metastable structure, in 
which half of the newly deposited Ge layer has interchanged 
position with the Sb surfactant layer. The configurations between
the initial and middle geometries and the middle and final geometries,
Fig.\ 3(b) and Fig.\ 3(d) respectively, correspond to the 
saddle point geometries which determine the activation energy
for the exchange. From our DFT/LDA calculations we found that the energy   
difference between structures 3(a) and 3(b) is 0.8 eV, and the 
energy difference between structures 3(c) and 3(d) is the same 
to within the accuracy of the results.  Similarly, the energy
difference between structures 3(c) and 3(b) and structures 3(e) and 3(d)
is 1.6 eV.  These two numbers correspond to the exchange activation
energy 
[0.8 eV, going from 3(a) to 3(c) through 3(b), or going from 
3(c) to 3(e) through 3(d)], 
and the de-exchange activation energy  
[1.6 eV, going from 3(c) to 3(a) through 3(b), or going from 
3(e) to 3(c) through 3(d)], for this hypothetical process.

We discuss next why these calculations give reasonable estimates
for the activation energies involved in 
surfactant mediated growth.  As far as the diffusion process
is concerned, it is typical for semiconductor surfaces to 
exhibit anisotropic diffusion constants depending on the surface
reconstruction, with the fast diffusion direction
along channels of atoms that are bonded strongly among themselves.
This is precisely the pathway we examined in Fig.\ 2.  
As far as the exchange process is concerned, it is 
believed that the only way in which atoms can exchange positions
in the bulk is through a concerted exchange type of motion,
as first proposed by Pandey for self diffusion in bulk Si
\cite{pandey86}.
This motion involves the breaking of the smallest possible
number of covalent bonds during the exchange, which 
keeps the activation energy relatively low. 
In the case of bulk Si, the activation energy for concerted
exchange is 4.5 eV \cite{pandey86}.  In the present case the activation
energy
is only 0.8 eV, because, unlike in bulk Si, the initial configuration
[Fig.\ 3(a)]
is not optimal, having the pentavalent Sb atoms as four-fold
coordinated (they would prefer three-fold coordination) and
the newly deposited Ge atoms as three-fold coordinated (they
would prefer four-fold coordination).  In the final configuration
[Fig.\ 3(e)], 
which has lower energy than the initial one, all atoms 
are coordinated properly (three-fold for Sb, four-fold for Ge).

While we have argued that the above described atomic processes
are physically plausible, we have not established neither their
uniqueness, nor their supremacy over other possible atomic motions.
In fact, the calculations of Schroeder et al.\ \cite{schroeder98}
discussed in section III.2,
are much more realistic as far as the exchange of single 
adatoms with surfactant atoms on terraces is concerned.
However, those calculations refer to a single event, and the 
formation of an additional substrate layer could (and probably does)
involve additional steps in the exchange process due to the
double-layer
nature of the Si(111) substrate.
In our calculations, the structure of the layer below the surfactant
is compatible with the lower half of the substrate double layer, 
so that the process of exchange can proceed with very similar steps
to complete the double-layer growth.
In this sense, we feel that the barriers we obtained are not too far
from realistic values. To keep our discussion general we will consider
the two sets of energy barriers as corresponding to a range of
physical systems: the first set is suggested by our results
($E_{d}=0.5$ eV, $E_{ex}=0.8$
eV, $E_{de-ex}=1.6$ eV) and the second by the 
results  of Schroeder et al.\ ($E_{d}=0.25$
eV, $E_{ex}=0.27$ eV, $E_{de-ex}=1.1$ eV).

\subsection*{3. De-exchange and generalized diffusion}

It is clear that diffusion and exchange processes affect the morphology
of the growing film. What about de-exchange processes? The energy
barrier for de-exchange is significantly larger than the other two
energy barriers. Are de-exchange events frequent enough to have any
effect on the growth mode? Or, to be more quantitative, suppose an
adatom exchanged with a surfactant atom; will it de-exchange before
another adatom exchanges in its vicinity? To answer this question we
assume that the time scale associated with a process with energy barrier
$E$ is $\nu^{-1}\exp(E/kT)$, where $\nu=10^{13}$ sec$^{-1}$ is the basic
attempt rate, $k$ is the Boltzmann constant and $T$ is the temperature.
Even at the fairly low temperature of $350^\circ$C and with the large
de-exchange barrier of 1.6 eV the time it takes an atom to de-exchange
is only 0.9 seconds. The time it takes to grow a layer at a typical flux
of 0.3 layers/minute is 200 seconds. Therefore an atom will de-exchange
quite a few times before it will interact with additional atoms in the
same layer. We conclude that de-exchange processes can influence the
growth mode and should not be ignored.

The above discussion changes our view of diffusion in
surfactant mediated epitaxy. The effect of diffusion cannot be simply
understood by comparing $E_{d}$ with $E_{ex}$, because after an
adatom has exchanged it may still continue to diffuse on top of the
surfactant layer by de-exchanging with a surfactant atom. It is
instructive to compare the effective diffusion constant, $D_{eff}$,
which corresponds to this complex diffusion process, with the bare
diffusion constant of an adatom on a surface (without surfactants),
$D=\nu a^2\exp(-E^{(b)}_{d}/kT)$. Here $a$ is the lattice constant
and $E^{(b)}_{d}$ is the energy barrier for bare diffusion.

To calculate $D_{eff}$, we consider the case $E_{ex}>E_{d}$ (a
similar calculation can be done for the opposite case and yields an
identical result). An effective diffusion hop consists of a de-exchange
event followed by several microscopic diffusion hops and finally an
exchange
event. We calculate $D_{eff}$ from the expression
$D_{eff}=A^2/\tau_{eff}$,
where $\tau_{eff}$ is the average time it takes to carry out an
effective
diffusion hop. $A$ is the average distance an atom travels during
such a hop, and obeys the relation $A=a\sqrt{n}$, where $n$ is the
average number of microscopic diffusion hops the atom carries out
between the
de-exchange and exchange events. $n$ is easily calculated as the ratio
between the time for an exchange event and the time for a microscopic
diffusion
hop. This leads to the result $n=\exp[(E_{ex}-E_{d})/kT]$.
$\tau_{eff}$ is the time it takes to carry out a de-exchange event
followed by an exchange event. Therefore,
$\tau_{eff}=\nu^{-1}[\exp(E_{de-ex}/kT)+\exp(E_{ex}/kT)]$. The final
expression for the effective diffusion constant is
\begin{equation}
D_{eff}=D~\frac{\exp[(E^{(b)}_{d}-E_{d})/kT]}{1+\exp[(E_{de-ex}-E_
{ex})/kT]}.
\label{effdiff}
\end{equation}
Clearly, a comparison of $E_{d}$ with $E_{ex}$ does not tell us
much about the magnitude of $D_{eff}$. Passivation of the surface by the
surfactant implies $E^{(b)}_{d}>E_{d}$ and $E_{de-ex}>E_{ex}$.
Thus both the numerator and the denominator are larger than 1, and the
question is which one is larger. For the values of energy barriers
calculated from first-principles (both ours and those of Schroeder et
al.) $E_{de-ex}-E_{ex}\approx 0.8$ eV. The denominator of Eq.\
(\ref{effdiff}) is therefore a very large number (between $10^3$ and
$10^6$ for typical temperatures). The numerator is much smaller and
therefore $D\gg D_{eff}$, i.e. diffusion is suppressed. This is not
necessarily the case for all surfactants. For example, for surfactants
which are less efficient in passivating the surface $E_{de-ex}-E_{ex}$
may be comparable or even smaller than $E^{(b)}_{d}-E_{d}$, which
would lead to diffusion enhancement. This may be the case for group-III
and group-IV surfactants, which enhance diffusion according to
experiments \cite{iwanari90,iwanari92a,wilk94,iwanari92b,hwang98}. It
would be interesting to
check this
possibility with DFT/LDA calculations. An interesting conclusion is
that effective diffusion can be suppressed by surfactants even if
$E_{ex}>E_{d}$; i.e., a surfactant can enhance diffusion on top of
the surfactant layer, and at the same time suppress effective diffusion,
which takes into account de-exchange processes.

\subsection*{4. Island-edge passivation}

At this stage, it is tempting to claim that we have reached a much
better understanding of the surfactant effect. However, as we show next,
this is not so. In fact, a more careful analysis shows that suppression
of diffusion has nothing to do with the explanation of the surfactant
effect, and that two surfactants, which lead to the same value of
$D_{eff}$ may induce very different growth modes. This happens because
different surfactants may vary drastically in their ability to
passivate steps or island edges. We will see that the issue of
island-edge passivation is crucially important to the morphology of the
growing film and its surface. It is especially important for the ability
of the surfactant to suppress 3D islanding in heteroepitaxy.

To understand the role of island edges in the determination of growth
modes, we have to understand the reason for the formation of 3D islands
in heteroepitaxial growth. They form because their presence facilitates
strain relaxation in a much
more efficient way than in flat layers. For example, Tersoff and Tromp
\cite{tersoff93} calculated the elastic energy per unit volume,
$E_{el}$, of a strained rectangular island of lateral dimensions $s$ and
$t$ (measured in units of the lattice constant). They showed that
\begin{equation}
E_{el}\sim\left(\frac{\ln s}{s}+
\frac{\ln t}{t}\right).
\label{elas}
\end{equation}
The energy of a narrow island is thus smaller than the energy of a wide
one. Therefore, after a monolayer-high island has grown beyond a certain
width, it is beneficial to grow another layer on top of it rather than
make it wider. The film then tends to grow in narrow and fairly tall 3D
islands. The kinetic process which prevents the island from growing
farther laterally is the detachment of atoms from the island edges. If
such detachment processes are suppressed, the island will not reach its
equilibrium shape. It will tend to be too wide and flat. It is quite
obvious that surfactants which passivate island edges will also suppress
detachment events. Hence, they may change the growth mode from 3D
islanding to layer-by-layer growth. Suppression of diffusion may not be
sufficient to suppress 3D islanding, since detachment of atoms from
island edges may lead to islanding even with very little diffusion.
Passivation of island edges, on the other hand, can change the growth
mode even if diffusion is not enhanced.

We now use our knowledge of the chemical nature of different surfactants
to speculate about their ability to passivate island edges: 
group-V atoms (especially As and Sb) should be effective in passivating
steps and island edges on the (111) and (100) surfaces of tetravalent
semiconductors such
as Si and Ge. This is because group-V atoms prefer to have three-fold
coordination, in which they form three strong covalent bonds with their
neighbors using three of their valence electrons, while the other two
valence electrons remain in a low-energy lone-pair state. This is
precisely what is needed for passivation of both terrace and step
geometries on the (111) and (100) surfaces of the diamond lattice, which
are characterized by three-fold coordinated atoms. On the other hand, it
is expected that elements with the same valence as the substrate, or
noble metals, will not be effective in passivating step edges. In the
case of the tetravalent semiconductors Si and Ge, for example, the
elements Sn and Pb have the same valence, and while they can form full
passivating layers on top of the substrate, they clearly cannot
passivate the step geometries since they have exactly the same valence
as the substrate atoms and hence can only form similar structures.  
Analogously, certain noble metals can form a passivating monolayer on
the semiconductor surface, but their lack of strong covalent bonding
cannot affect the step structure. We note that not all noble metals
behave in a similar manner, with some of them forming complex structures
in which they intermix with the surface atoms of the substrate
(such as Ag on the Si(111) surface), in which case it is doubtful that
they will exhibit good surfactant behavior.

\subsection*{5. Kinetic Monte Carlo simulations}

We have given a plausibility argument that surfactants suppress 3D
islanding in heteroepitaxy by limiting atom detachment from island edges
and not by suppressing diffusion. The complexity of the growth process
does not allow us to give a more rigorous argument. However, our ansatz
can be tested quite easily by carrying out kinetic Monte Carlo (KMC)
simulations of homoepitaxial and heteroepitaxial growth, in which all
the relevant microscopic processes occur randomly with rates determined
by the corresponding activation energies. 
Accordingly, we consider a system in which the processes examined above 
are operative, and the activation energies corresponding to
them are the ones obtained from the DFT/LDA calculations
for the hypothetical cases illustrated in Fig.\ 2 and Fig.\ 3.

For simplicity, our simulation was carried out on a cubic lattice. Atoms 
land on the surfactant covered surface with a flux of 0.3 layers/second
(a typical value of the flux in experiments), and diffuse on top of the
surfactant. They can exchange with surfactant atoms and become buried
underneath the surfactant layer. A buried atom can de-exchange with a
surfactant atom and float on top of the surfactant layer again. This can
happen provided the buried atom does not have lateral bonds with other
atoms underneath the surfactant layer. If it is bonded laterally, we
consider this atom as being part on an island edge. An atom attached to
an island underneath
the surfactant layer can detach from the island edge and float on top of
the surfactant layer. This detachment process is of major importance, as
discussed above. However, it involves breaking of lateral bonds between
the detaching atom and the island edge. This will be taken into account
in
the activation barrier for detachment. Also, we did not allow
simultaneous breaking of two or more lateral bonds, so an atom attached
to an island edge by more than a single lateral bond cannot detach. 
A diffusing atom can attach to a step or an island edge. The activation
barriers for attachment and detachment processes depend on whether the
surfactant passivates steps and island edges or not. Barriers for
detachment from an island edge also depend on whether the island is
strained or not.

We now describe the results of the simulations that we did under various
conditions. In each case we give a detailed list of the activation
energy values. First, we studied homoepitaxial growth, i.e., we
considered a system without lattice mismatch and hence no strain
effects. We investigated
the influence of island-edge passivation (IEP) on surface morphology. To
that end we carried out KMC simulations of a surface of
size $100\times 100$ at a temperature of 600$^\circ$C, and deposited on
it 0.15
of a layer. The values of the activation energies used were $E_{d}=0.5$
eV, $E_{ex}=0.8$ eV and $E_{de-ex}=1.6$ eV. The energy barrier for
detachment from an island edge (provided only one lateral bond is
broken) was $E_{det}=3$ eV for a surfactant which passivates
island-edges and $E_{det}=1.6$ eV for a surfactant that does not.
Typical morphologies are shown in Fig.\ 4(a), with IEP and Fig.\ 4(b)
without IEP.  
Evidently, there is a marked
difference between the growth process with and without IEP. First,
surfactants that passivate island edges lead to a significantly higher
island density in submonolayer growth. Secondly, with IEP the island
edges are very rough, while without passivation the islands are
faceted. As discussed above, experimental results indicate
\cite{voigtlander95} that surfactants that suppress 3D islanding also
increase the island density in homoepitaxy. The rough island edges are
also observed experimentally. This gives strong support to the IEP
ansatz. We note that the high density of islands induced by surfactants
which passivate island edges is not a result of suppression of
diffusion. It arises from the fact that adatoms can cross passivated
island edges without attaching to them and then nucleate on a flat part
of the surface, thus generating more islands.

In Figs.\ 4(c) and 4(d) we present results from similar simulations
with another set of activation barriers:
$E_{d}=0.5$ eV, $E_{ex}=0.3$ eV, $E_{de-ex}=1.1$ eV, $E_{det}=2.5$ with
IEP and $E_{det}=1.6$ without. Although these barriers are very
different from the ones we used to produce Figs.\ 4(a) and 4(b), the
morphologies are very similar. The change in the energy barriers has not
influenced the island densities, nor has it affected the shape of the
islands significantly. The only noticeable effect is that the  shape of
the islands in Fig.\ 4(c) is more fractal-like than the shape of the
ones in Fig.\ 4(d). Note that in the first set of energy barriers
$E_{d}<E_{ex}$, whereas in the second one the opposite holds. Thus the
relation between $E_{d}$ and $E_{ex}$ does not have a significant
influence on
the growth morphology. Based on these results, we expect the energy
barriers we calculated using DFT/LDA
and those of Schroeder et al.\ (see above) to lead to similar growth
morphologies; i.e. the difference between these two sets of activation
barriers is irrelevant for the determination of the growth mode.
The results of Fig.\ 4 support our generalized diffusion analysis. The
two sets of energy barriers we used give the same value for the
effective diffusion constant, $D_{eff}$, according to Eq.\
(\ref{effdiff}). This is the reason the final surface morphologies are
so similar. 

It is also important 
to check the temperature dependence of the growth process
in the case of surfactants with IEP. To this end we performed KMC
simulations of a $300\times 300$ lattice with activation energies
identical to
the ones used for Fig.\ 4(a). The resulting surface morphologies after
deposition of 0.15 of a layer are shown in Fig.\ 5 for three
different temperatures: 600$^\circ$C, 700$^\circ$C and 850$^\circ$C.
At all three temperatures IEP leads to a high density of compact islands
with rough edges. The island density decreases with temperature. All of
these observations are consistent with experimental results
\cite{voigtlander95}. 

Finally, we consider the effects of strain in surfactant 
mediated heteroepitaxial growth. Strain is difficult to include in an
atomistic calculation in a self consistent manner. Here we will rely on 
the theory developed by Tersoff and Tromp \cite{tersoff93} for the
elastic energy of strained islands on a substrate (see Eq.\
(\ref{elas})). In analogy with this theory, we postulate that the effect
of strain is to alter 
the strength of the bonds in elastically strained islands
according to the expression of Eq.\ (\ref{elas}), which 
depends on the island size through the values of $s$ and $t$.
The most important consequence of this effect is a change in the  
activation energy for detachment of atoms from island edges, $E_{det}$,
since this process involves breaking of a lateral bond which is strongly
affected by strain. $E_{det}$ 
will now depend on the island size. The other barriers,
having to do with processes that take place on top of the 
surfactant (diffusion and exchange on terraces and island edges),
will be unaffected to lowest order by the presence of strain.
Therefore, the only important change in the kinetics comes
from an island-size dependent detachment rate, given by
\begin{equation}
E_{det}=\epsilon_0 + \epsilon_1\left(\frac{\ln s}{s}+
\frac{\ln t}{t}\right),
\label{de-ex}
\end{equation}
where $\epsilon_0=E_{de-ex}$ for surfactants which passivate island
edges, and $\epsilon_0=0$ when there is no IEP.

In our simulations we take the value $\epsilon_1 = 3.0$ eV,
which is a reasonable number for the typical strength of 
bonds and the amount of strain involved in the
systems of interest (4\% for the case of Ge on Si). 
As was done in the case of homoepitaxial growth, 
we first study the effects of passivation of island
edges on surface morphology. We simulated a system of size $100\times
100$ at a
temperature of 300$^\circ$C, and deposited on it one layer.
The values of the activation energies used were $E_{d}=0.5$
eV, $E_{ex}=0.8$ eV and $E_{de-ex}=1.6$ eV. The results are shown in
Figs.\ 6(a) and 6(b) for the cases with and without IEP, respectively.
Different surface heights are represented by different colors; white is
the initially flat substrate. 
The system without IEP shows clear 3D islanding [Fig.\ 6(b)]. Most of
the substrate is exposed even after deposition of a full layer, and the
deposited material is assembled in faceted tall (up to 7 layers high)
islands. The surfactant which passivates island edges, on the other
hand, suppressed 3D islanding completely [Fig.\ 6(a)]. Most of the
surface is covered by one layer (the blue color), with some small
one-layer-high islands and holes in it.

We have also checked the influence of changes in the values of the
activation barriers, by repeating the simulations with $E_{d}=0.5$
eV, $E_{ex}=0.3$ eV and $E_{de-ex}=1.1$ eV. The results are presented in
Figs.\ 6(c) and 6(d) for the cases with and without IEP, respectively.
The system without IEP does not show any change. In the case with IEP
the densities of islands and holes decreased and their sizes increased
accordingly. But the growth mode remained layer-by-layer, and 3D
islanding was entirely suppressed. 
These results together with the results on homoepitaxy demonstrate
convincingly that island-edge passivation, and not suppression of
diffusion, is responsible for the surfactant effect.

To study the effect of temperature on the growth mode in heteroepitaxial
growth with IEP, we simulated a system of size $300\times 300$, with the
activation barriers $E_{d}=0.5$ eV, $E_{ex}=0.8$ eV and $E_{de-ex}=1.6$
eV. The resulting morphologies at the temperatures of 300$^\circ$C,
350$^\circ$C, 400$^\circ$C and 450$^\circ$C are shown in Fig.\ 7.
In the first two cases, growth is essentially indistinguishable
from the case of homoepitaxy, with a high 
density of small islands. However, at $T=400^\circ$C, despite the 
small rise of only 50$^0$C,
a dramatically different growth mode is evident, with a large 
number of tall 3D islands and a substantial amount of the 
substrate left uncovered. This trend is even more evident
at the higher temperature of 450$^\circ$C. We also carried out
simulations of heteroepitaxial growth on vicinal surfaces, with exactly
the same parameters as those of Fig.\ 7, but starting from a system with
atomic steps present on the substrate. Fig.\ 8 shows the results of
these KMC simulations for the same temperatures as in Fig.\ 7. Again,
the surfactant suppressed 3D islanding at low temperatures, but not at
high temperatures. This is precisely the type of abrupt transition from
layer-by-layer growth at low temperature, to 3D island growth 
at higher temperature observed experimentally for the strained
heteroepitaxial systems, such as Ge/Si with Sb as a surfactant.

\section*{V. Discussion}

We have provided a critical review of the literature on
surfactant mediated semiconductor epitaxy with emphasis on comparisons
between experimental observations and model calculations. Our main goal
was to arrive at a consistent explanation of the mechanism by which
surfactants suppress 3D islanding in heteroepitaxial growth.

There is a vast number of experimental articles on the subject and we
gathered most of them in tables according to the relevant combination of
deposit-surfactant-substrate materials. The most important message,
which one can take from these experimental studies, is that in
semiconductor epitaxy surfactants can be divided into two categories. In
the first category we have surfactants which lead to an anomalously high
island density in submonolayer homoepitaxy, and also suppress 3D
islanding in heteroepitaxy. The second category consists of surfactants
which lead to step flow growth in homoepitaxy and are inefficient in
suppressing 3D islanding in heteroepitaxy.

Explanations of the surfactant effect have focused on the relation
between the activation energy for adatom diffusion on top of the
surfactant layer and the barrier for exchange of an adatom with a
surfactant atom. The rationale was that surfactants of the first
category
suppress 3D islanding and increase the island density because they
suppress diffusion. Suppression of diffusion was associated with
relatively easy exchange processes. Surfactants of the second category,
on the other hand, are thought to enhance diffusion, and exchange
processes were expected to be relatively difficult.

Several first-principles calculations of the diffusion and exchange
barriers have been carried out for various systems. In a typical
calculation, specific paths for diffusion and exchange were proposed and
total energies of the system in relaxed configurations along these paths
were calculated. This allows fairly accurate estimates of the relevant
energy barriers. Different studies arrived at different conclusions
about the barriers mainly because the paths proposed were different.
Thus the main deficiency of these microscopic calculations is their
inability to predict the correct kinetic path for the process under
consideration. 

We proposed a new scenario for the explanation of the surfactant effect.
According to our ansatz, neither the relation between diffusion and
exchange nor suppression of diffusion are relevant for the explanation
of the surfactant effect. Instead, we argued that the efficiency of a
surfactant is determined by its ability to passivate island edges.
Surfactants which passivate island edges also lead to an anomalously
high density of islands and suppress 3D islanding. We supplied ample
evidence for this scenario. The most convincing evidence comes from
kinetic Monte Carlo simulations of the growth process and from the
comparison of the results with experimental observations. Using
realistic activation energies we showed that a surfactant that
suppresses diffusion, but does not passivate island edges, does not
suppress 3D islanding. It also does not lead to a very high density of
islands. Moreover, the islands generated in the growth process mediated
by such a surfactant are faceted and do not have the rough edges
observed in experiments. By contrast, island-edge passivation does lead
to suppression of 3D islanding and to islands of a shape consistent 
with the experimentally
observed shape. The temperature dependence of the island density, as
well
as the abrupt transition from layer-by-layer to 3D growth as the
temperature is raised, are predicted correctly by simulations with IEP.

The evidence we provided for the validity of our scenario, although
convincing, is far from being a rigorous proof. In fact, it is based on
a very simplified model, which fails to take into account various
aspects of the experimental system that may be important. For example,
the use of an isotropic square lattice is not appropriate for all cases 
(the substrates with diamond lattice and (111) orientation have a 
hexagonal surface lattice, whereas those with (100) orientation 
have a square lattice but 
exhibit strong anisotropy in the directions parallel and perpendicular
to 
the surface dimers);
nor do we account for the fact that in some cases the film grows in
bilayers (as in (111) substrates) rather than in monolayers
(as in (100) substrates). We do not properly treat the issue of
the critical island size, which may be very large in Si homoepitaxy or
in heteroepitaxy of Ge on Si. There is therefore room for further
discussion and more detailed modeling of surfactant mediated thin film
growth. 

There are various unresolved issues in surfactant mediated epitaxial
growth, which we have not discussed. Perhaps the most important among
them is the issue of strain relaxation. Heteroepitaxial films grown in
the layer-by-layer growth mode are initially highly strained. This
strain energy must somehow relax after growth of a few layers. Indeed,
dislocations appear in the film during surfactant mediated
heteroepitaxy. In some cases these dislocations do not thread the film
and hence do not harm its epitaxial quality, but in other cases they do.
It is therefore very important to study strain relaxation in the films.
Some experimental studies have been carried out, but their description
is beyond the scope of the present article. To the best of our
knowledge, there has not been any detailed 
theoretical work on the problem. 

Another important issue is related to the fact that inevitably
some of the surfactant layer gets trapped in the growing film.
This leads to unintended doping if the surfactant is not isoelectronic
with the deposited material.  This could be beneficial, if high
levels of doping are desired, or detrimental, if a film of high 
purity is desired.  In any case, controlling the amount of 
the incorporated surfactant by carefully adjusting external conditions
(such as flux rate, temperature, surface preparation) is highly 
desirable \cite{xie96}.  A better understanding of the surfactant
effect,
along the lines proposed here for the DDP model,
will probably go a long way toward controlling 
the electronic properties of the film, which are strongly
influenced by surfactant incorporation, strain relaxation defects
and surface morphology. 
Further
research in these directions is necessary and essential before
surfactant mediated growth can become useful in practical applications. 

This work was supported by the Office of Naval Research Grant
\# N00014-95-1-0350, and by THE ISRAELI SCIENCE FOUNDATION founded by
The Israeli Academy of Sciences and Humanities. D.K.\ is the incumbent 
of the Ruth Epstein Recu Career Development Chair.

\newpage

\newpage
\centerline{Figure Captions}

{\bf Figure 1:} Schematic illustration of important mechanisms
in surfactant mediated growth on a substrate (represented 
by white circles) with a full monolayer surfactant coverage
(represented by continuous shaded area): (a) diffusion on 
terraces and steps for surfactant that
passivates step edges; (b) exchange at terraces and passivated steps;
(c) de-exchange at terraces and passivated steps; 
(d) diffusion on terrace and exchange at non-passivated steps;
(e) de-exchange at terrace and at non-passivated steps.

{\bf Figure 2:} Representative surface diffusion pathway, top and side
views.  
The dark circles represent the substrate atoms, the light circles
the surfactant atoms.  The smaller gray circle represents an 
extra atom deposited on top of the surfactant layer, at different
positions.  
The geometries correspond to a Ge adatom on a Si(111) surface (the
substrate)
covered by a monolayer of Sb (the surfactant) in a ($2\times1$) chain
reconstruction.

{\bf Figure 3:} Representative exchange pathway.  The color scheme is
the
same as in Fig. 2.  (a) Structure with one layer of newly deposited
atoms on top of the surfactant layer.  The geometries depicted in  
(b), (c), (d) are the intermediate
structures during a concerted exchange that brinks the surfactant
layer on top of the newly deposited layer, shown as the final 
configuration in (e).  Structure (c) is metastable, while structures
(b) and (d) are saddle-point configurations.
Solid lines linking the atoms correspond to covalent bonds, while 
dashed lines correspond to borken bonds.  The geometries correspond to 
the same physical system as in Fig. 2.

{\bf Figure 4:} Kinetic Monte Carlo simulations of surfactant mediated
homoepitaxy in the DDP model, 
on a substrate of size $100\times 100$ at a temperature of
600$^\circ$C. A total of 0.15 monolayer of new material has been
deposited: 
(a) Simulations with IEP, with  the activation energies
$E_{d}=0.5$ eV, $E_{ex}=0.8$ eV, $E_{de-ex}=1.6$ eV and $E_{det}=3$ eV.
(b) Simulations without IEP, with  the activation energies $E_{d}=0.5$
eV, $E_{ex}=0.8$ eV, $E_{de-ex}=1.6$ eV and $E_{det}=1.6$ eV. 
(c) Simulations with IEP, with  the activation energies $E_{d}=0.5$ eV,
$E_{ex}=0.3$ eV, $E_{de-ex}=1.1$ eV and $E_{det}=2.5$ eV. 
(d) Simulations without IEP, with  the activation energies $E_{d}=0.5$
eV,
$E_{ex}=0.3$ eV, $E_{de-ex}=1.1$ eV and $E_{det}=1.6$ eV. IEP clearly
increases the island density and significantly affects the island shape.

{\bf Figure 5:} Kinetic Monte Carlo simulations of homoepitaxial
surfactant
mediated growth in the DDP model, 
with IEP on a substrate of size $300\times 300$. The
activation
energies were $E_{d}=0.5$ eV, $E_{ex}=0.8$ eV, $E_{de-ex}=1.6$ eV and
$E_{det}=3$ eV. A total of 0.15 monolayer of new material has been 
deposited at 600$^\circ$C, 700$^\circ$C and 850$^\circ$C.
The high
density of small islands at low temperature is evident, as well as 
the decrease of the island density with increasing temperature.

{\bf Figure 6:} Kinetic Monte Carlo simulations of surfactant mediated
heteroepitaxy in the DDP model, 
on a substrate of size $100\times 100$ at a temperature of
300$^\circ$C. A total of one monolayer of new material has been
deposited. 
(a) Simulations with IEP, with the activation energies
$E_{d}=0.5$ eV, $E_{ex}=0.8$ eV and $E_{de-ex}=1.6$ eV. 
(b) Simulations
without IEP, with  the activation energies $E_{d}=0.5$ eV, $E_{ex}=0.8$
eV and $E_{de-ex}=1.6$ eV. 
(c) Simulations with IEP, with  the
activation energies $E_{d}=0.5$ eV, $E_{ex}=0.3$ eV, $E_{de-ex}=1.1$ eV.
(d) Simulations without IEP, with the activation energies $E_{d}=0.5$
eV, $E_{ex}=0.3$ eV and $E_{de-ex}=1.1$. Different colors indicate
different surface heights. The surfactant which passivates island edges
suppresses 3D islanding completely and induces layer-by-layer growth.
Without IEP 3D islands form on the film. They reach a height of 7 layers
after deposition of one layer of material. 

{\bf Figure 7:} Kinetic Monte Carlo simulations of heteroepitaxial
surfactant
mediated growth in the DDP model, 
on a substrate of size $300\times 300$ with IEP. The
activation
energies were $E_{d}=0.5$ eV, $E_{ex}=0.8$ eV and $E_{de-ex}=1.6$ eV. A
total of one monolayer of new material has been 
deposited at 300$^\circ$C, 350$^\circ$C, 400$^\circ$C and
450$^\circ$C.  
The different colors indicate surface heights.
The transition between layer-by-layer growth and 3D island growth takes
place somewhere between 350$^\circ$C and 400$^\circ$C.

{\bf Figure 8:} Kinetic Monte Carlo simulations of heteroepitaxial
surfactant
mediated growth in the DDP model, 
on a {\em vicinal} substrate of size $300\times 300$
with IEP.
The activation
energies were $E_{d}=0.5$ eV, $E_{ex}=0.8$ eV and $E_{de-ex}=1.6$ eV. A
total of one monolayer of new material has been 
deposited at 300$^\circ$C, 350$^\circ$C, 400$^\circ$C and
450$^\circ$C.  
The different colors indicate surface heights.

\end{document}